\documentclass[conference]{IEEEtran}

\usepackage{array}
\usepackage[utf8]{inputenc}
\usepackage{amsmath, commath}
\usepackage{amsthm}
\usepackage{amsfonts}
\usepackage{amssymb}
\usepackage{bm}
\usepackage{bbm}
\usepackage{graphicx}
\usepackage{graphics}
\usepackage{xcolor}
\usepackage{float}
\usepackage{tikz}

\usepackage{epstopdf}
\usepackage[export]{adjustbox}
\usepackage[list=true]{subcaption}
\usepackage{color}
\usepackage{algorithm}
\usepackage{algorithmic}

\usepackage[justification=centering]{caption}
\usepackage{enumerate} 
\usepackage{cite}
\usepackage{suffix}
\usepackage{mathtools}
\usepackage{tabularx, booktabs}
\usepackage{multirow}
\usepackage{diagbox}
\usepackage{amsfonts}
\usepackage{amssymb}
\usepackage{textcomp}
\usepackage{dsfont} 
\usepackage{svg}
\usepackage{pgfplots,pgfplotstable}
\pgfplotsset{compat=1.3}
\usetikzlibrary{automata, positioning, arrows,shapes.multipart,fit}

\def\0{{\mathbf 0}}

\def\i1{{\mathbf 1}}

\usepackage{cite}
\usepackage{amsmath,amssymb,amsfonts}
\usepackage{algorithmic}
\usepackage{graphicx}
\usepackage{subcaption}
\usepackage{textcomp}
\usepackage{url}

\usepackage{array}
\usepackage{stfloats}

\usepackage{multicol}

\usepackage{multirow} %to pros8esa gia tous pinakes sta result
\def\BibTeX{{\rm B\kern-.05em{\sc i\kern-.025em b}\kern-.08em
    T\kern-.1667em\lower.7ex\hbox{E}\kern-.125emX}}

\title{ClusterSlice: A Zero-touch Deployment Platform for the Edge Cloud Continuum}
\author{Lefteris Mamatas (Member, IEEE), Sotiris Skaperas (Member, IEEE),  Ilias Sakellariou (Member, IEEE)\\
Department of Applied Informatics, University of Macedonia, GR-54636 Thessaloniki, Greece\\
\{emamatas, sotskap, iliass\}@uom.edu.gr\\
\thanks{This research was partially funded by the Greek Ministry of Education and Religious Affairs for the project "Enhancing Research and optimizing University of Macedonia's administrative operation.}
\thanks{Corresponding Author: Sotiris Skaperas (email: \textit{sotskap@uom.edu.gr}).}
}

\begin{document}
\maketitle
\begin{abstract}
We demonstrate ClusterSlice, an open-source solution for automated Kubernetes-center deployments for the edge continuum. ClusterSlice is an infrastructure-as-a-service, platform-as-a-service, and application-as-a-service solution, supporting: (i) declarative deployment slice definitions; (ii) infrastructure-on-demand capabilities over multiple heterogeneous domains; (iii) composable Kubernetes deployments, supporting multi-clustering as well as various Kubernetes flavors and intra-cluster/inter-cluster network plugins; (iv) configurable application deployment; and (v) experimentation automation. 
\end{abstract}

\maketitle

%\section{Plan}

%Answered RQ
%\begin{itemize}
%    \item Heterogeneity via abstractions
%    \item zero touch, declarative
%    \item Scalability
%    \item Reproducability of Experiments
%    \item 
%\end{itemize}

%Enabler for studying a plethora of problems as in
%\begin{itemize}
%    \item Resource Allocation
%    \item Intend based Orchestration
%    \item Energy Efficiency
%    \item security, 
%\end{itemize}
%
%Description of Github Experiment + Video

\section{Introduction}

ClusterSlice is an open-source Kubernetes-centered Edge Cloud solution, able to convert compute resources from bare-metal or hypervisor-level to fully-operatable cloud-network slices. One of its stronger features is the introduction of well-designed abstractions, that reduce deployment complexity with improved reliability and reproducibility, configuring the corresponding slice parts in a parallel fashion and zero-touch manner. ClusterSlice is cloud-native and fully integrated to Kubernetes, a widely used orchestration facility with impressive automation, fault-tolerance, scalability, resource optimization capabilities and, most importantly for the context of this work, a highly modular architecture. The latter allows Clusterslice to implement its functionality by harvesting on the concepts of Custom Resource Definitions (CRDs) and the Operator pattern, and to be fully managed via the {\tt kubectl} tool. Due to this tight integration, it benefits from all Kubernetes reliability features. ClusterSlice supports:
\begin{itemize}
\item {\it Declarative automated operation}: 
Slice definition is done at a higher level in the form of YAML files, and includes compute resources to be utilized, Kubernetes configuration, and the application modules to be installed. 

\item {\it Infrastructure-as-a-service capabilities}: ClusterSlice supports the utilization of heterogeneous physical and virtual resources through a common abstraction, including open testbed infrastructures (e.g., CloudLab), as well as Virtual Machines (VMs) in XCP-ng and VirtualBox virtualization systems. 

\item {\it Platform-as-a-service features}: It supports multiple Kubernetes flavors, such as vanilla Kubernetes (k8s), k0s, k3s, and microk8s, as well as network plugins for both intra-cluster (e.g., Flannel, Calico, Kube-ovn, etc.) and inter-cluster communication (e.g., Submariner).
\item {\it Application-as-a-service attributes}: The slice definition supports deployment of configurable application modules, i.e., the definition of applications to deploy, k8s extensions and modular OS configurations, ensuring maximum compatibility with heterogeneous systems. 
\item {\it Multi-cluster and multi-domain capabilities}: ClusterSlice operation can span across multiple heterogeneous deployment environments through technology-specific infrastructure managers, establishing multi-cluster operation and communication, such as Liqo, OCM, or Submariner. 
\item {\it Experimentation automation}: ClusterSlice is enables advanced, automated deployment and management of experiments, based on its characteristics of reproducability in the deployment of slices, and actively supported automation capabilities, such as comparable evaluations of network plugins and anomaly detection workflows. 
\end{itemize}

ClusterSlice provides infrastructure-, platform- and application-as-a-service capabilities over the edge continuum. Relevant infrastructure control solutions include SLICES (https://portal.slices-sc.eu) and FABRIC \cite{fabric} test-beds, which enable multiple infrastructure federations to inter-operate (e.g., Emulab, CloudLab, and others). Their novel resource control features and APIs (e.g., jFed, python APIs, AMs, etc.) make them re-usable for higher-level deployment abstractions. For example, CloudNativeLab (https://practicum.cloudnativelab.ilabt.imec.be) enables the creation of a Kubernetes cluster on SLICES testbeds, using a defined resource configuration for the cluster. Furthermore, 5G-CDN \cite{5g-cdn} and NECOS \cite{necos2} are slicing infrastructures capable of allocating resources across multi-domain infrastructures.

The platform- or application-oriented solutions, include EdgeNet \cite{edgenet1}, SLATE \cite{Slate_2018} and Kubernetes-native testbed  (https://github.com/kubernetes-native-testbed). EdgeNet \cite{edgenet1} is a platform-as-a-service solution over multiple, crowd-sourced k8s worker nodes, implementing an edge continuum through CRDs and operators. SLATE federates science platforms, allowing sites to delegate service deployment and configuration to designated application administrators. The Kubernetes-native testbed integrates various microservices-based applications, CI/CD environments, and monitoring tools. 

ClusterSlice supports multiple infrastructure managers under a common design abstraction, so all above mentioned solutions could be jointly utilized and federated. Furthermore, it also supports both platform-as-a-service capabilities (e.g., composable Kubernetes clusters or multi-cluster deployments) and application-as-a-service capabilities, defining the application or Kubernetes extensions to use: all at the level of a declarative slice definition.

In the rest of sections, a brief presentation of ClusterSlice is covered in Section \ref{sec:clusterslice-description}, whereas sections \ref{sec:demo-description} and \ref{sec:conclusions} detail the demo run and conclude the paper, respectively.

\section{Proposed System}
\label{sec:clusterslice-description}
%It supports multiple physical and virtual infrastructure controllers (i.e., infrastructure managers in ClusterSlice terminology).
%The latter is achieved by harvesting on the power of {\it bash} scripting, SSH and Ansible (https://www.ansible.com), thus eliminating any dependencies on external libraries or APIs. 
The automated transformation of resources from bare-metal or hypervisor setups into fully-operational Kubernetes-based edge cloud deployments, follows a unique approach: it benefits from k8s capabilities based on novel design abstractions. 

ClusterSlice design is empowered by the Kubernetes paradigm, and its implementation heavily utilizes Custom Resource Definitions and Kubernetes Operators, thus inheriting the {\it reliability, scalability}, and {\it resource optimization} capabilities of Kubernetes. The implementation carefully avoids the use of APIs or technology-specific software, relying only on common tools such as {\it bash}, command-line utilities, and SSH through Ansible. This preserves {\it simplicity, portability}, and {\it composability} in line with the Unix philosophy.

Furthermore, ClusterSlice introduces {\it innovative abstractions} on different levels of its architecture: 
\begin{itemize}
    \item {\it node and cluster level automation abstractions}, i.e., Resource Managers (RMs) and Slice Operators (SOs), respectively. These act like digital twins of compute resources (i.e., physical hosts or VMs) and k8s clusters. ClusterSlice allocates and declaratively configures such containers, which, in turn, represent and actively configure the resources they represent. 
    \item {\it  infrastructure manager abstractions} that interface with diverse cloud/virtualization systems (e.g., VirtualBox or XCP-ng) and even testbed facilities (e.g., CloudLab), maximizing {\it compatibility} across heterogeneous systems and seamless {\it future-proof} integration.
    \item {\it multi-cluster and multi-domain abstractions}, to create and oversee multiple clusters and deploy the respective software, e.g., Liqo, Karmada or OCM, in the form of application modules.
\end{itemize}

A more detailed description of the ClusterSlice architecture and its individual components can be found in \cite{clusterslice-tr-2023}. One of ClusterSlice's strongest features lies in its ability to deploy a diverse range of characteristics for the deployment slice through a declarative definition. This is further demonstrated in the following section. 

\begin{figure*}
    \setlength{\columnsep}{0.1em} % Adjust the column separation here
    \begin{multicols}{2}
        \centering
        \small
        \begin{verbatim}
apiVersion: "swn.uom.gr/v1"
kind: MultiClusterSliceRequest
metadata:
  name: liqo
spec:
  name: liqo
  namespace: swn
  deploymentstrategy: balanced
  credentials:
    username: clusterslice
    password: sha-512-encoded-password
  clusters:
    - name: liqo
      deploymentdomain: swntestbed
      infrastructure:
        masters:
          count: 1
          osimage: "ubuntu-22-clean"
          mastertype: "vm"
        workers:
          count: 1
          osimage: "ubuntu-22-clean"
          workertype: "vm"
      kubernetes:
        kubernetestype: "vanilla"
        networkfabric: "flannel"
      applications:
        - name: liqo-master
          scope: cluster
          parameters:"{peers:[liqo1,liqo2]}"
    - name: liqo1
      deploymentdomain: lefteris
      infrastructure:
        masters:
          count: 1
          osimage: "ubuntu-22-clean"
          mastertype: "vm"
        workers:
          count: 1
          osimage: "ubuntu-22-clean"
          workertype: "vm"
      kubernetes:
        kubernetestype: "vanilla"
        networkfabric: "flannel"
      applications:
        - name: liqo-peer
          scope: cluster
          sharefile: "liqo1-peer-join.sh" 
    - name: liqo2
      deploymentdomain: cloudlab
      infrastructure:
        masters:
          count: 1
          osimage: "UBUNTU22-64-STD"
          osaccount: "lmamatas"
          mastertype: "pc3000"
        workers:
          count: 1
          osimage: "UBUNTU22-64-STD"
          osaccount: "lmamatas"
          workertype: "pc3000"
      kubernetes:
        kubernetestype: "vanilla"
        networkfabric: "flannel"
      applications:
        - name: liqo-peer
          scope: cluster
          sharefile: "liqo2-peer-join.sh"
        \end{verbatim}
    \end{multicols}
    \caption{A Multi-Cluster \& Multi-Domain Slice Descriptor}
    \label{fig:example-definition}
    \vspace{-4mm}
\end{figure*}

\section{Demo Description}
\label{sec:demo-description}

\begin{figure}
  \centering
  \includegraphics[width=0.48\textwidth,trim=0 5 0 1, clip]{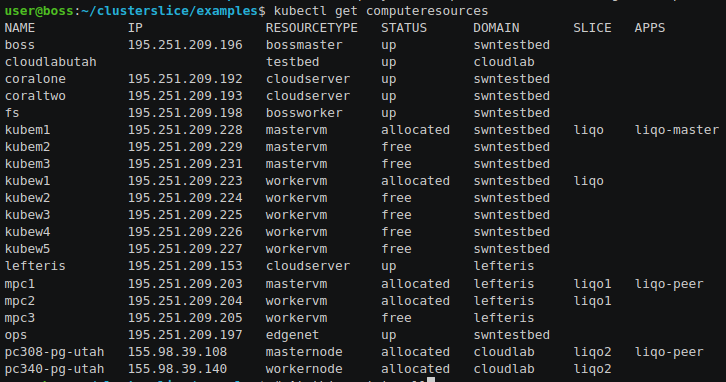}
  \caption{Compute Resources after the Slice Deployment}
  \label{fig:clusterslice-example}
  \vspace{-3mm}
\end{figure}
The demo involves the preparation of various slice deployment descriptors (e.g., as specified by the demo audience), in the form of YAML files. Subsequently, these descriptors are used for deploying corresponding cloud-network slices. Here, we provide a specific example involving three Kubernetes clusters across heterogeneous domains, accompanied by a multi-cluster management system, specifically Liqo. The scenario considers the following diverse set of deployment environments:
 \begin{itemize}
     \item A local two-server test-bed hosting the XCP-ng cloud environment, known as ``swntestbed''.
     \item A workstation PC hosting VirtualBox, named ``lefteris''.
     \item The CloudLab test-bed, spread across various locations in the USA, referred to as ``cloudlab''.
\end{itemize}

The declarative definition of the deployment is illustrated in Fig. \ref{fig:example-definition}.  
In the YAML file, we present the overarching properties of the experimentation slice, i.e., the name, namespace, server deployment strategy, and user credentials. After that, the specification of the three clusters follows. Each one of these specifications includes information on node, Kubernetes, and application requirements:
\begin{itemize}
  \item {\it Cluster node requirements} such as the number of nodes in each cluster, the OS image to be deployed in master and worker nodes, and the node types. It should be noted that each cluster possesses its own deployment domain and unique resource type, for instance {\tt pc3000} physical server at the CloudLab cluster or VM in the XCP-ng or VrtualBox clusters.
  \item {\it Kubernetes cluster configuration} supporting the full range of versions, i.e., vanilla, k0s, k3s, etc., and multiple network fabrics, e.g., Flannel, Calico, KubeRouter, etc., allowing for any (valid) combination of the previous, include requesting a specific version of the above, enhancing the reproducability aspect of the deployment. 
   \item Application requirements, that for this scenario include the Liqo Kubernetes extension. It is interesting to note that the Liqo master deployment (cluster named {\tt liqo}), specifies, in a rather declarative manner, the participating clusters in the multi-cluster deployment (\tt {'peers': '[liqo1,liqo2]'}).
\end{itemize}

The deployment is initiated by issuing a single {\tt kubectl apply} command. The first phase involves allocation of resources for the three clusters, which is handled by the appropriate infrastructure managers depending on their type; for instance in the case of CloudLab by the \textit{Testbed Infrastructure Manager} whereas in the case of VMs, an embedding process might also occur which leads to VMs allocated to specific cloud servers, via the \textit{Cloud Infrastructure Managers}. 

During the second phase, the deployment process involves several steps, including: (a) OS image installation and configuration, (b) Kubernetes cluster setup for the three distinct clusters (liqo, liqo1, and liqo2), and (c) the establishment of the Liqo multi-cluster management system, in this specific scenario.

After completing the above phase, Liqo is successfully implemented across three diverse clusters, as illustrated in Fig.\ref{fig:clusterslice-example}. At any deployment stage, executing the same command provides the status of all involved compute resources. Multiple such examples will be available during the demo.

\section{Conclusions}
\label{sec:conclusions}

ClusterSlice offers a declarative, effortless, zero-touch solution for transforming testbed resources, even from bare-metal, into fully operational Kubernetes slices. This is achieved via well-designed abstractions and heavy exploitation of the Kubernetes CRD and Operator patterns. 

%It should be noted that 
The software is available as open source via its GitHub repository (https://github.com/SWNRG/clusterslice), that also includes a video of the demo presented above. We have run multiple experimentation scenarios, as presented in \cite{clusterslice-tr-2023}, deploying a total of $410$ clusters and $1530$ nodes, without a single failure, an initial strong indication of the reliability advantages of ClusterSlice. 

\section*{Acknowledgments}
This research  is part of the project ``Experimental Assessment Of Network And Compute Resource Optimization Methods (NCOPT)'', that has received funding from University of Macedonia Research Fund.

\bibliography{clusterslice.bib}
\bibliographystyle{IEEEtran}

\end{document}